\documentstyle[12pt,epsf]{article}
\begin{document}
\baselineskip=18pt
\begin{titlepage}
\begin{flushright}
  KUNS-1653\\[-1mm]
  hep-ph/0003231
\end{flushright}

\begin{center}
  \vspace*{1.4cm}
  
  {\large\bf Novel Relations between Lepton and Quark Mixings}
  \vspace{1cm}
  
  Masako~Bando\footnote{E-mail address: bando@aichi-u.ac.jp},
  Taichiro~Kugo\footnote{E-mail address: 
    kugo@gauge.scphys.kyoto-u.ac.jp} and
  Koichi~Yoshioka\footnote{E-mail address:
    yoshioka@gauge.scphys.kyoto-u.ac.jp}
  \vspace{5mm}
  
  $^*$ {\it Aichi University, Aichi 470-0296, Japan}\\
  $^{\dagger,\ddagger}$ {\it Department of Physics, Kyoto University
    Kyoto 606-8502, Japan}
  \vspace{1.5cm}
  
  \begin{abstract}
    A novel relation is found which gives the 2-3 lepton mixing angle
    $\theta_{\mu\tau}$ in terms of quark masses and CKM mixing:
    $$\tan\theta_{\mu\tau}=(m_b/m_s)V_{cb}.$$
    This relation, which is remarkably in good agreement with the
    quark data and the large lepton mixing recently observed, is a
    kind of $SO(10)$ GUT relation similar to the celebrated bottom-tau
    mass ratio in $SU(5)$. The GUT models in which this relation hold
    should have a `twisted $SO(10)$' structure between the second and
    third generations, in order to really explain the large lepton
    mixing, or equivalently, the right-hand side of the relation.
  \end{abstract}
\end{center}
\end{titlepage}

One of the most remarkable facts in recent particle physics is the
very large lepton mixing observed in SuperKamiokande~\cite{SK}, which
has revealed a sharp contrast to the quark sector where the CKM
mixings are all small. Why is such a difference possible between the
quark and lepton sectors? This is really a challenge to the attempts
for grand unified theories (GUTs). Clearly any GUTs which treat the
three generations of quarks and leptons as a mere repetition no longer
work. We need some structural changes between the generations and a
number of models accounting for the large lepton mixing angles have
recently been proposed~\cite{model}.

In our recent study of a supersymmetric $E_6$ grand unified
model~\cite{e6}, we found a novel relation
\begin{equation}
  \tan\theta_{\mu\tau}\;=\;{m_b\over m_s}V_{cb},
  \label{novel}
\end{equation}
or equivalently,
\begin{eqnarray}
  \sin^2 2\theta_{\mu\tau} &=& \frac{4V_{cb}^2
    \left(\displaystyle\frac{m_s}{m_b}\right)^2}{\left[V_{cb}^2 
      +\left(\displaystyle\frac{m_s}{m_b}\right)^2\right]^2}\,.
  \label{sumrule}
\end{eqnarray}
All the quantities here are those measured at the GUT scale. Although
this relation was found in our explicit $E_6$ GUT model, it actually
does not depend on the details of the model. On the contrary, it turns
out to have a much more generality (for
example,~\cite{e6,so10,ABB}). It is a kind of an $SO(10)$ GUT relation
similar to the famous $SU(5)$ relation $m_b/m_\tau=1$~\cite{btau}, as
the derivation shows which we give now.

This relation (\ref{novel}) results if the theory gives the following
mass matrices for the up-quark, down-quark and charged-lepton sectors:
\begin{eqnarray}
  \bar u_LM_u u_R &=& \mu_u\ \bar u_L\pmatrix{
    O(\lambda^4) & f\lambda^2 \cr
    O(\lambda^2) & 1     \cr}u_R, \nonumber\\[1ex]
  \bar d_LM_d d_R &=&\mu_d\ \bar d_L\pmatrix{
    e\lambda^2 & f\lambda^2  \cr
    h & 1 \cr}d_R, \nonumber\\[1ex]
  \bar e_RM_e^{\rm T}e_L &=&\mu_d\ \bar e_R\pmatrix{
    O(\lambda^2) & O(\lambda^2)  \cr
     h & 1 \cr}e_L,
   \label{mass}
\end{eqnarray}
where we have written $2\times2$ matrices concentrating our attention
only to the second and third generations, for simplicity, and we have
normalized the 2-2 entries to be 1 by factoring out the mass scale
parameters $\mu_u$ and $\mu_d$. $\lambda$ denotes a number of the
order of the Cabibbo angle $\sin\theta_{\rm C}\sim 0.22$, and all the
coefficients, $e,\,f,\,h$, are assumed to be of order $\lambda^0$. The
point here is that (i) the second columns of $M_u$ and $M_d$ are
commonly given by $(f\lambda^2,\, 1)^{\rm T}$, and (ii) the second rows
of $M_d$ and $M_e^{\rm T}$ commonly by $(h,\, 1)$. The first property
(i), being a proportionality condition only of the second columns of
$M_u$ and $M_d$, will clearly hold if, as a sufficient condition, the
third generation up and down quarks fall into 
a {\it single representation} of the GUT group $SO(10)$ or larger. The
second property (ii) is satisfied if the right-handed down quark and
the left-handed charged lepton in each generation are contained in a
single ${\bf 5^*}$ of $SU(5)$. Indeed the equality of the 2-2 entries of
$M_d$ and $M_e^{\rm T}$ is the same as the famous $m_b=m_\tau$
relation, and we are now demanding the same $SU(5)$ relation to hold
also for the 2-1 entries, $h$.

Given these forms of the mass matrices in any case, we can find the
following unitary matrices diagonalizing the matrices
$M_iM_i^\dagger$, $\,U_i(M_iM_i^\dagger)U_i^\dagger=$ diagonal 
for $i=u,\ d,\ e$: 
\begin{eqnarray}
  U_u &=& \pmatrix{
    1 & -f\lambda^2 \cr
    f\lambda^2 & 1  \cr }, \nonumber\\[1ex]
  U_d &=& \pmatrix{
    1 & \displaystyle -{f+eh\over 1+h^2}\lambda^2 \cr
    \displaystyle {f+eh\over 1+h^2}\lambda^2 & 1 \cr},\nonumber\\[1ex]
  U_e &=& \pmatrix{
    \displaystyle {1\over \sqrt{1+h^2}} &
    \displaystyle -{h\over \sqrt{1+h^2}} \cr
    \displaystyle {h\over \sqrt{1+h^2}} &
    \displaystyle {1\over \sqrt{1+h^2}} \cr},
  \label{U}
\end{eqnarray}
up to smaller corrections suppressed by a factor $\lambda^2$. Note
that the down-quark matrix $M_d$ itself is diagonalized as
\begin{eqnarray}
  && U_dM_dU_e^\dagger \,=\, \pmatrix{
    m_s & 0 \cr 0 & m_b \cr }, \nonumber\\[1ex]
  && m_s \,=\, \mu_d\,{e-fh\over \sqrt{1+h^2}}\lambda^2, 
  \qquad m_b \,=\, \mu_d\sqrt{1+h^2}
  \label{down}
\end{eqnarray}
by using $U_d$ and the {\it charged-lepton}'s $U_e$. This is due to
the $SU(5)$ property (ii) mentioned above. Now, using Eqs.~(\ref{U})
and (\ref{down}), we can calculate the element $V_{cb}$ of the CKM
matrix  $V=U_uU_d^\dagger$ and immediately find a relation
\begin{equation}
  V_{cb} \;=\; {m_s\over m_b}h + O(\lambda^4).
  \label{vcb}
\end{equation}
This is almost the relation we desire. However, to reveal the $SO(10)$
property (i) working behind, let us re-derive this in another way.

For this purpose, we switch to the basis in which the up quark mass
matrix is diagonalized, then the quark doublets are transformed 
as $(\bar u_L,\,\bar d_L) \to (\bar u_L,\,\bar d_L) U_u$ so that the
down-quark mass matrix turns into
\begin{equation}
  M'_d \,=\, U_uM_d  \,=\,
  \pmatrix{V_{cs} & V_{cb} \cr V_{ts} & V_{tb} \cr}
  \pmatrix{m_s & 0 \cr 0 & m_b \cr}
  \pmatrix{U_{\mu2} & U_{\mu3} \cr U_{\tau2} & U_{\tau3} \cr},
  \label{UMd}
\end{equation}
where $V_{ij}$ are the matrix elements of the CKM matrix $V$,
and $U_{\mu i}$ and $U_{\tau i}$ are those of the charged-lepton
$U_e$. In this up-quark-diagonal basis, the 1-2 entry $M'_{d\,12}$
becomes {\it zero} up to $O(\lambda^4)$. This important fact comes
from the $SO(10)$ property (i) in the above; indeed, the left
multiplication of $U_u$ (together with the right multiplication of a
suitable matrix) makes the up-quark mass matrix $M_u$ diagonal and so,
taking also account of the $\lambda$-hierarchical structure of $M_u$,
we see that the 1-2 entry of $U_uM_u$ vanishes up to an $O(\lambda^4)$
correction. But the $M_u$ and $M_d$ have the second columns exactly
proportional to each other so that $U_uM_d$ should also have vanishing
1-2 entry up to $O(\lambda^4)$. Computing the 1-2 matrix 
element $M'_{d\,12}$ from the expression (\ref{UMd}), we obtain 
\begin{equation}
  V_{cs}m_sU_{\mu3}+V_{cb}m_bU_{\tau3} \,=\, 0+O(\lambda^4).
\end{equation}
or equivalently, using $V_{cs}=1+O(\lambda^2)$,
\begin{equation}
  -{U_{\mu3}\over U_{\tau3}} \;=\; {m_b\over m_s}V_{cb} +
  O(\lambda^2).
  \label{qed}
\end{equation}
This is just the same relation as Eq.~(\ref{vcb}).

This Eq.~(\ref{qed}) is nothing but the relation (\ref{novel})
announced at the beginning, provided that the MNS lepton mixing comes
almost solely from the charged-lepton sector. The MNS matrix is
generally given by $V_{\rm MNS}=U_eU_\nu^\dagger$ in terms of the
above charged-lepton's $U_e$ and the neutrino's $U_\nu$ which
diagonalizes the light neutrino Majorana mass matrix $M_\nu$ 
(as $U_\nu^*M_\nu U_\nu^\dagger=$ diagonal). So, the condition is that 
(iii) the neutrino diagonalization matrix $U_\nu$ is essentially a
unit matrix:
\begin{equation}
  U_\nu \,=\, 1+O(\lambda^2).
  \label{Unu}
\end{equation}
This holds, for instance if the neutrino Majorana mass matrix $M_\nu$
takes the form of the hierarchical structure
\begin{equation}
  M_\nu \,\propto\, \pmatrix{\lambda^{2\alpha} & \lambda^\alpha\cr
    \lambda^\alpha& 1\cr},
  \label{hier}
\end{equation}
in the fermion basis in which $M_u,\,M_d$ and $M_e$ take the forms
(\ref{mass}). If the power $\alpha$ equals 2, $M_\nu$ has the
same structure as the up-quark mass matrix $M_u$ in Eq.~(\ref{mass})
and the Eq.~(\ref{Unu}) results. Such a parallelism between $M_\nu$
and $M_u$ seems natural to appear in $SO(10)$ GUT models, although
being not automatic from the symmetry alone, of course, since it
depends on the structure of the right-handed neutrino Majorana mass
matrix. A hierarchical structure (\ref{hier}) of $M_\nu$ comes out,
for instance, in a supersymmetric $E_6$ GUT model studied by the
present authors recently.

We have thus clarified how and under which conditions our relation
(\ref{novel}) can be derived. It is actually remarkable that the
relation (\ref{novel}), or the equivalent 
one (\ref{sumrule}), {\it is} in good agreement with the observed
large lepton mixing angle $\theta_{\mu\tau}$ and the experimental data
of quark masses and mixing. In Fig.~1, we show the value 
of $\sin^2 2\theta_{\mu\tau}$ which the relation (\ref{sumrule})
predicts for various values of the quark mass ratio $m_s/m_b$ and CKM
matrix element $V_{cb}$ within their experimental errors.

We have now come to another important point. The relation
(\ref{novel}) itself results from the conditions (i)--(iii) alone and
is well satisfied by the experimental values. However, for the model
satisfying these three conditions to really predict the large lepton
mixing angle $\theta_{\mu\tau}$, or equivalently to reproduce the
experimental value for the right-hand side (quark side) of the
relation (\ref{novel}), it must realize the relation $h\simeq 1$. This
is clear since the lepton mixing angle is given 
by $\tan\theta_{\mu\tau}=h$ as is seen from $U_e$ in Eq.~(\ref{U}).

The requirement $h\simeq 1$ gives a very non-trivial condition to the
models. To see the point it is better to use explicit models realizing
the above relations. We here consider two such 
models, `generation flipped' $SO(10)$ model by Nomura and Yanagida
(NY)~\cite{so10}, and `$E$-twisted' $E_6$ model~\cite{e6} by the
present authors. As far as the second and third generations are
concerned, these two models are essentially the same (although the
latter model is more predictive, and superior at the point that it can
additionally explain the largeness of $m_t/m_b$). In these models, the
required hierarchical structure by powers of $\lambda$ of the mass
matrices $M_u$, $M_d$ and $M_e$ in Eq.~(\ref{mass}) can be realized by
the Froggatt-Nielsen mechanism~\cite{FN}, and the powers of $\lambda$
are determined by the $U(1)$ quantum numbers assigned to the
fermions. Thus the fermions belonging to the same multiplet must have
the same $U(1)$ quantum numbers and hence the same powers of $\lambda$
in their masses. The mass terms at issue are
\begin{equation}
  \mu_d\,\bar d_{L3}\,(\,h\, d_{R2}+ 1\,d_{R3}\,)
\end{equation}
for down quarks, which should be compared with
\begin{equation}
  \mu_u\,\bar u_{L3}\,(\,O(\lambda^2)\, u_{R2}+ 1\,u_{R3}\,)
\end{equation}
for up quarks, where the numbers in the suffices denote the
generation. The ratio $\lambda^2:1$ between $u_{R2}$ and $u_{R3}$ for
the latter is different from $h:1$ between $d_{R2}$ and $d_{R3}$ which
should be order 1 here. In the NY model the third generation fermions
belong to a single multiplet {\bf 16} of the GUT group $SO(10)$ (the
condition (i)). So, in order to have this different powers 
of $\lambda$ (the condition (ii)), the second generation up 
quark $u_{R2}$ must belong to a different multiplet from that of down
quark $d_{R2}$. This is actually the case in the NY model, where the
up-quark $u_{R2}\in SU(5)\,{\bf 10}$ belong to a {\bf 16} as usual
while the down-quark $d_{R2}\in SU(5)\,{\bf 5}^*$ come from another
multiplet {\bf 10} of $SO(10)$. {\it It is this mechanism of violating
  the parallel generation structure that is required for explaining
  the observed large lepton mixing}.

It is interesting to see how the same is realized in the $E_6$ model
of ours. It is always possible to find an $SO(10)$ group under which 
a {\bf 10} and a ${\bf 5}^*$ of $SU(5)$ are combined into 
a {\bf 16}. Therefore, although in the NY model the third generation
fermions are in a {\bf 16} while the second generation {\bf 10} 
and ${\bf 5}^*$ of $SU(5)$ come separately from {\bf 16} and {\bf 10}
of $SO(10)$, the converse is also true; namely, if we take another
$SO(10)$ group, which may be called `flipped $SO(10)$', then the
second generation fermions fall into a single {\bf 16} while the third
generation fermions are separated into ${\bf 16}$ and ${\bf 10}$.
(In the case of $E_6$, moreover, the two $SO(10)$ groups, the original
one and the flipped one, are mutually converted into each other by 
an $SU(2)_E$ rotation with angle $\pi$ contained in $E_6$, hence
explaining the name `$E$-twisting'.) This is exactly what happens in
our $E_6$ model for the $SO(10)$ subgroup, which is chosen by the
first step spontaneous breaking $E_6\to SO(10)$; among three
generations, each belonging to a {\bf 27} of $E_6$ in the beginning,
the second and first generation fermions each fall into a {\bf 16}
under the $SO(10)$ subgroup after the breaking, while the third
generation splits into $SU(5)\; {\bf 10}\subset{\bf 16}$ 
and $SU(5)\; {\bf 5}^*\subset{\bf 10}$. 

We suspect that the observed very large mixing in the lepton/neutrino
sector indicates the existence of such a `flipping' or `twisting'
structure as an essential ingredient of the GUT models. Remarkably,
the $E_6$ model already prepares such a mechanism intrinsically; the
fundamental representation {\bf 27} contains {\bf 16} and {\bf 10} 
of $SO(10)$ and so two $SU(5)\;{\bf 5}^*$s. They can be mixed or even
twisted completely by the internal rotation in $E_6$.

Finally we add a comment that another relation can be obtained from
the same conditions if we include the first generation also. Then, the
mass matrices become $3\times3$, and the previous condition (i) is
generalized to the proportionality of the third column of $M_u$ 
and $M_d$ while the second property (ii) for $M_d$ and $M_e$ and (iii)
are kept the same. Then the same argument as before leads to a
relation
\begin{equation}
  \tan\theta_{\mu\tau} \;=\; \frac{V_{ub}}{V_{us}}\frac{m_b}{m_s}.
\end{equation}
This also roughly agrees with the experimental data, although the
inclusion of the first generation data becomes less reliable
generally.

To conclude, we have found in this letter the novel relations between
the quark and lepton mixing angles in GUTs. The relations hold under
the simple conditions, especially, if the third generation up and down
quarks come from a single multiplet. It clearly requires $SO(10)$ or
larger unified gauge groups. We have also shown that the conditions
are naturally realized with the `twisted' generation structure of the
second and third generations. The experimental agreement of the
presented relations may corroborate these interesting GUT structures.

\subsection*{Acknowledgments}

We would like to thank to Y.~Nomura, N.~Okamura and T.~Yanagida for
stimulating discussions. We also thank the Summer Institute 99
held at Yamanashi, Japan. M.~B., T.~K.\ and K.~Y.\ are supported in
part by the Grants-in-Aid for Scientific Research No.~09640375,
No.~10640261, and the Grant-in-Aid No.~9161, respectively, from the
Ministry of Education, Science, Sports and Culture, Japan.

\newpage
\vspace*{3cm}

\begin{figure}[htbp]
  \begin{center}
    \leavevmode
    \epsfxsize=11cm \ \epsfbox{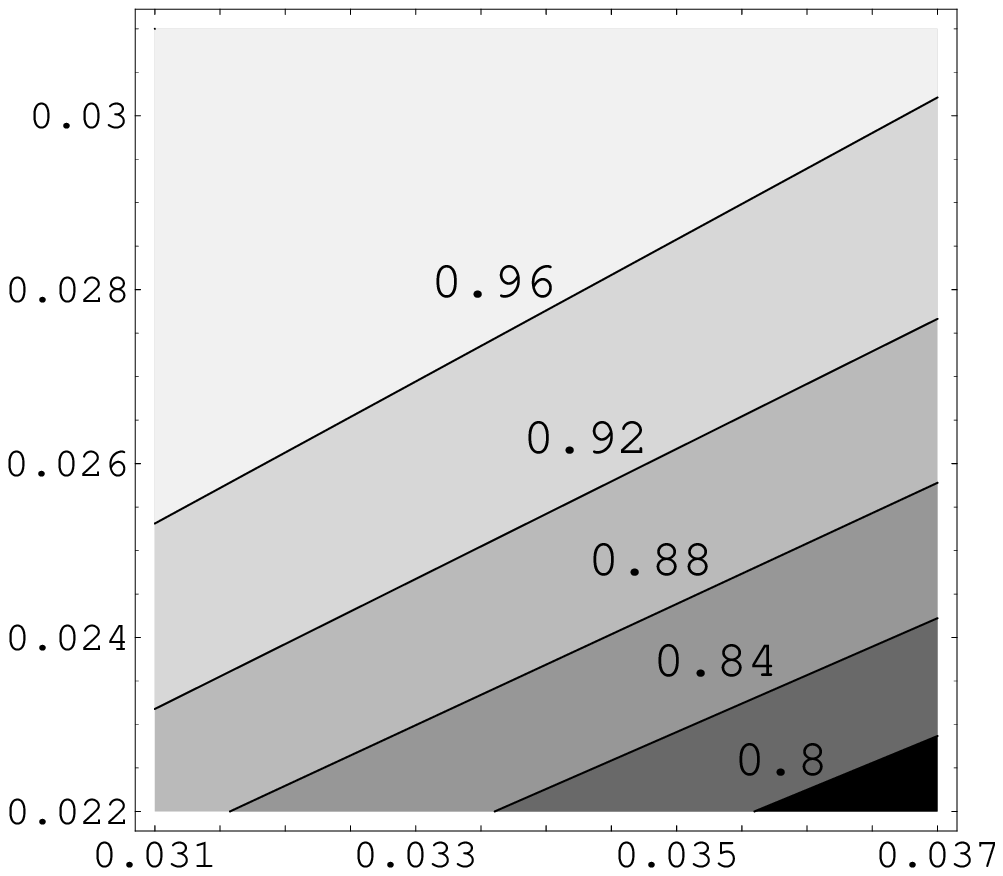}
    \put(-287,285){$m_s/m_b$}
    \put(7,16){$V_{cb}$}
    \caption{The prediction of the lepton 2-3 mixing angle 
      $\sin^2 2\theta_{\mu\tau}$ from the relation
      (\ref{sumrule}). The square parameter region is the experimental
      uncertainties of $m_s/m_b$ and $V_{cb}$. In almost all range,
      the relation is consistent with the observations.}
  \end{center}
\end{figure}

\end{document}